\documentclass[twocolumn,showpacs,preprintnumbers,amsmath,amssymb,prl]{revtex4}

\usepackage{graphicx}
\usepackage{dcolumn}

\usepackage{mathptmx, courier, pifont}
\usepackage[scaled=0.92]{helvet}
\usepackage[T1]{fontenc}
\usepackage{textcomp}

\usepackage{bm}

\begin{document}

\title{ Projected shell model description for nuclear isomers }

\author{Yang Sun$^{1,2}$}

\affiliation{$^1$Department of Physics, Shanghai Jiao Tong
University, Shanghai 200240, P. R. China\\$^2$Joint Institute for
Nuclear Astrophysics, University of Notre Dame, Notre Dame, Indiana
46545, USA}

\begin{abstract}
Nuclear isomer is a current research focus. To describe isomers, we
present a method based on the Projected Shell Model. Two kinds of
isomers, $K$-isomers and shape isomers, are discussed. For the
$K$-isomer treatment, $K$-mixing is properly implemented in the
model. It is found however that in order to describe the strong
$K$-violation more efficiently, it may be necessary to further
introduce triaxiality into the shell model basis. To treat shape
isomers, a scheme is outlined which allows mixing those
configurations belonging to different shapes.
\end{abstract}


\maketitle


\section{Introduction}

A nuclear isomer is an excited state, in which a combination of
nuclear structure effects inhibits its decay and endows the isomeric
state with a lifetime that can be much longer than most nuclear
states. Known isomers in nuclei span the entire range of lifetimes
from 10$^{15}$ years for $^{180m}$Ta -- longer than the accepted age
of the universe -- to an informal rule of thumb on the lower side of
approximately 1 ns. Nuclear isomers decay predominantly by
electromagnetic processes ($\gamma$-decay or internal conversion).
There are also known instances of the decay being initiated by the
strong interaction ($\alpha$-emission) or the weak interaction
($\beta$-decay or electron capture). Decay by proton or neutron
emission, or even by nuclear fission, is possible for some isomers
(see recent examples \cite{Walker06,94Ag1,94Ag2}).

Often discussed in the literature are three mechanisms
\cite{Walker99} leading to nuclear isomerism, although new types of
isomer may be possible in exotic nuclei \cite{SM1}. It is difficult
for an isomeric state to change its shape to match the states to
which it is decaying, or to change its spin, or to change its spin
orientation relative to an axis of symmetry. These correspond to
shape isomers, spin traps, and $K$-isomers, respectively. In any of
these cases, decay to the ground state is strongly hindered, either
by an energy barrier or by the selection rules of transition.
Therefore, isomer lifetimes can be remarkably long. To mention a few
examples, an $I^\pi = 0^+$ excited state in $^{72}$Kr has been found
as a shape isomer \cite{Kr72}, a 12$^+$ state in $^{98}$Cd has been
understood as a spin trap \cite{Cd98}, and in $^{178}$Hf, there is a
famous 16$^+$, 31-year $K$-isomer \cite{Hf178}, which has become a
discussion focus because of the proposal of using this isomer as
energy storage \cite{Walker2}.

Detailed nuclear structure studies are at the heart of understanding
the formation of nuclear isomers with applications to many aspects
in nuclear physics. The study is particularly interesting and
important for unstable nuclei, such as those in neutron-rich,
proton-rich, and superheavy mass regions. In a quantum system, the
ground state is usually more stable than the excited states.
However, the lifetime of ground state of unstable nuclei is short,
which makes the laboratory study extremely difficult. In contrast,
nuclear isomers in those nuclei may be relatively easy to access
experimentally. Furthermore, the physics may be changed due to the
existence of isomers in those unstable nuclei. It has been pointed
out by Xu {\it et al.} \cite{Xu04} that in superheavy nuclei, the
isomeric states decrease the probability for both fission and
$\alpha$-decay, resulting in enhanced stability for these nuclei.
One expects that the isomers in very heavy nuclei could serve as
stepping stones toward understanding the single-particle structure
beyond the $Z=82$ and $N=126$ shell closures, which is the key to
locating the anticipated `island of stability' \cite{No254}.

Moreover, nuclear isomers may play a significant role in determining
the abundances of the elements in the universe \cite{Ani05}. In hot
astrophysical environments, an isomeric state can communicate with
its ground state through thermal excitations. This could alter
significantly the elemental abundances produced in nucleosynthesis.
The communication between the ground state of $^{26}$Al and the
first excited isomeric state in this nucleus has the consequence
that the astrophysical half-life for $^{26}$Al can be much shorter
than the laboratory value \cite{Al26}. One is just beginning to look
at the impact that nuclear isomers have on various other
nucleosynthesis processes such as the rapid proton capture process
thought to take place on the accretion disks of binary neutron
stars. There are cases in which an isomer of sufficiently long
lifetime (probably longer than microseconds) may change the paths of
reactions taking place and lead to a different set of elemental
abundances \cite{Wiescher05}.

With rapidly growing interest in the isomer study and increasing
possibility of experimental access to isomeric states, theoretical
effort is much needed. The present paper discusses a Projected Shell
Model (PSM) description for nuclear isomers. As isomeric states are
a special set of nuclear states, special emphasis is given when
these states are treated. In section II of the paper, we present a
description for $K$-isomers, in which $K$-mixing is emphasized. We
point out, however, that an extended PSM based on
triaxially-deformed basis is required to describe the strong
$K$-violation. In section III, shape isomer examples are presented
and a perspective how configurations with different shapes can be
mixed is outlined. Finally, the paper is summarized in section IV.

\section{$K$-mixing in the projected shell model}

Many long-lived, highly-excited isomers in deformed nuclei owe their
existence to the approximate conservation of the $K$ quantum number.
The selection rule for an electromagnetic transition would require
that the multipolarity of the decay radiation, $\lambda$, be at
least as large as the change in the $K$-value ($\lambda \ge \Delta
K$).  However, symmetry-breaking processes make possible transitions
that violate the $K$-selection rule.  A microscopic description of
$K$-violation is through the so-called $K$-mixing in the
configuration space. A theoretical model that can treat $K$-mixing
has preferably the basis states that are eigenstates of angular
momentum $I$ but labeled by $K$. Diagonalization of two-body
interactions mixes these states and the resulting wavefunctions
contain the information on the degree of $K$-mixing. In this kind of
approach, the mixing and its consequences are discussed in the
laboratory frame rather than in a body-fixed frame in which $K$ is
originally defined.

\subsection{The model}

The projected shell model (PSM) \cite{PSM,PSMCode} seems to fulfil
the requirement. It is a shell model that starts from a deformed
basis. In the PSM, the shell-model basis is constructed by
considering a few quasiparticle (qp) orbitals near the Fermi
surfaces and performing angular momentum projection (if necessary,
also particle-number projection) on the chosen configurations.  With
projected multi-qp states as the basis states of the model, the PSM
is designed to describe the rotational bands built upon qp
excitations.

Suppose that a PSM calculation begins with axially deformed Nilsson
single-particle states, with pairing correlations incorporated into
these states by a BCS calculation. This defines a set of deformed qp
states (with $a^\dagger_\nu$ and $a^\dagger_\pi$ being the creation
operator for neutrons and protons, respectively) with respect to the
qp vacuum $|0\rangle$. The PSM basis is then constructed in the
multi-qp states with the following forms
\begin{eqnarray}
{\rm e-e ~nuclei:} & & \{ |0\rangle, a^\dagger_\nu a^\dagger_\nu
|0\rangle, a^\dagger_\pi a^\dagger_\pi |0\rangle, a^\dagger_\nu
a^\dagger_\nu a^\dagger_\pi a^\dagger_\pi |0\rangle, \nonumber\\ & &
a^\dagger_\nu a^\dagger_\nu a^\dagger_\nu a^\dagger_\nu |0\rangle,
a^\dagger_\pi
a^\dagger_\pi a^\dagger_\pi a^\dagger_\pi |0\rangle, \ldots \}\nonumber\\
{\rm o-o ~nuclei:} & & \{ a^\dagger_\nu a^\dagger_\pi |0\rangle,
a^\dagger_\nu a^\dagger_\nu a^\dagger_\nu a^\dagger_\pi |0\rangle,
a^\dagger_\nu
a^\dagger_\pi a^\dagger_\pi a^\dagger_\pi |0\rangle, \nonumber\\
& & a^\dagger_\nu a^\dagger_\nu a^\dagger_\nu a^\dagger_\pi
a^\dagger_\pi a^\dagger_\pi |0\rangle, \ldots \}\nonumber\\
{\rm odd-\nu ~nuclei:} & & \{ a^\dagger_\nu |0\rangle, a^\dagger_\nu
a^\dagger_\nu a^\dagger_\nu |0\rangle, a^\dagger_\nu a^\dagger_\pi
a^\dagger_\pi |0\rangle, \nonumber\\ & & a^\dagger_\nu a^\dagger_\nu
a^\dagger_\nu a^\dagger_\pi a^\dagger_\pi |0\rangle, \ldots \}\nonumber\\
{\rm odd-\pi ~nuclei:} & & \{ a^\dagger_\pi |0\rangle, a^\dagger_\nu
a^\dagger_\nu a^\dagger_\pi |0\rangle, a^\dagger_\pi a^\dagger_\pi
a^\dagger_\pi |0\rangle, \nonumber\\ & & a^\dagger_\nu a^\dagger_\nu
a^\dagger_\pi a^\dagger_\pi a^\dagger_\pi |0\rangle, \ldots \}
\nonumber
\end{eqnarray}
The omitted index for each creation operator contains labels for the
Nilsson orbitals.  In fact, this is the usual way of building
multi-qp states \cite{Jain95,Soloviev,Zeng02,Xu04}.

The angular-momentum-projected multi-qp states, each being labeled
by a $K$ quantum number, are thus the building blocks in the PSM
wavefunction, which can be generally written as
\begin{equation}
|\psi^{I,\sigma}_M\rangle=\sum_{\kappa,K\le I}
f^{I,\sigma}_{\kappa} \hat P^{\,I}_{MK}|\phi_\kappa\rangle =
\sum_{\kappa} f^{I,\sigma}_{\kappa} \hat
P^{\,I}_{MK_\kappa}|\phi_\kappa\rangle .
\label{wavef}
\end{equation}
The index $\sigma$ denotes states with same angular momentum and
$\kappa$ labels the basis states.  $\hat P^{\,I}_{MK}$ is the
angular-momentum-projection operator \cite{PSM} and the coefficients
$f^{I,\sigma}_{\kappa}$ are weights of the basis states. The weights
$f^{I,\sigma}_{\kappa}$ are determined by diagonalization of the
Hamiltonian in the projected spaces, which leads to the eigenvalue
equation (for a given $I$)
\begin{equation}
\sum_{\kappa^\prime}\left(H_{\kappa\kappa^\prime}-E_\sigma
N_{\kappa\kappa^\prime}\right) f^\sigma_{\kappa^\prime} = 0.
\label{eigen}
\end{equation}
The Hamiltonian and the norm matrix elements in Eq. (\ref{eigen})
are given by
\begin{equation}
H_{\kappa\kappa^\prime}=\langle\phi_\kappa | \hat H \hat
P^I_{K_\kappa K^\prime_{\kappa^\prime}} | \phi_{\kappa^\prime}
\rangle , ~~~~~~~~~~ N_{\kappa\kappa^\prime}=\langle\phi_\kappa |
\hat P^I_{K_\kappa K^\prime_{\kappa^\prime}} |
\phi_{\kappa^\prime} \rangle .
\label{z}
\end{equation}
Angular-momentum-projection on a multi-qp state
$|\phi_\kappa\rangle$ with a sequence of $I$ generates a band. One
may define the rotational energy of a band (band energy) using the
expectation values of the Hamiltonian with respect to the
projected $|\phi_\kappa\rangle$
\begin{equation}
E^I_\kappa={H_{\kappa\kappa}\over
N_{\kappa\kappa}}={{\langle\phi_\kappa | \hat H \hat P^I_{K_\kappa
K_\kappa} | \phi_\kappa \rangle}\over {\langle\phi_\kappa | \hat
P^I_{K_\kappa K_\kappa} | \phi_\kappa \rangle}} .
\label{bande}
\end{equation}

In a usual approximation with independent quasiparticle motion, the
energy for a multi-qp state is simply taken as the sum of those of
single quasiparticles.  This is the dominant term.  The present
theory modifies this quantity in the following two steps. First, the
band energy defined in Eq. ({\ref{bande}) introduces the correction
brought by angular momentum projection and the two-body
interactions, which accounts for the couplings between the rotating
body and the quasiparticles in a quantum-mechanical way. Second, the
corresponding rotational states (labeled by $K$) are mixed in the
subsequent procedure of solving the eigenvalue equation
({\ref{eigen}). The energies are thus further modified by
configuration mixing.

For deformed states with axial symmetry, each of the basis states in
(\ref{wavef}), i.e. the projected $|\phi_\kappa\rangle$, is a
$K$-state. For example, an $n$-qp configuration gives rise to a
multiplet of $2^{n-1}$ states, with the total $K$ expressed by $K =
|K_1 \pm K_2 \pm \cdots \pm K_n|$, where $K_i$ is for an individual
neutron or proton.  In this case, shell model diagonalization, i.e.
solving the eigenvalue equation (\ref{eigen}), is equivalent to
$K$-mixing. The degree of $K$-mixing can be read from the resulting
wavefunctions.

The above discussion is independent of the choice of the two-body
interactions in the Hamiltonian.  In practical calculations, the PSM
uses the separable form of Hamiltonian with pairing plus
quadrupole-quadrupole terms (these have been known to be essential
in nuclear structure calculations \cite{Zuker96}), with inclusion of
the quadrupole-pairing term
\begin{equation}
\hat H = \hat H_0 - {1 \over 2} \chi \sum_\mu \hat Q^\dagger_\mu
\hat Q^{}_\mu - G_M \hat P^\dagger \hat P - G_Q \sum_\mu \hat
P^\dagger_\mu\hat P^{}_\mu .
\label{hamham}
\end{equation}
The strength of the quadrupole-quadrupole force $\chi$ is determined
in such a way that it holds a self-consistent relation with the
quadrupole deformation $\varepsilon_2$.  The monopole-pairing force
constants $G_M$ are
\begin{equation}
\begin{array}{c}
G_M = \left[ G_1 \mp G_2 \frac{N-Z}{A}\right] ~A^{-1} ,
\label{GMONO}
\end{array}
\end{equation}
with ``$-$" for neutrons and ``$+$" for protons, which roughly
reproduces the observed odd--even mass differences in a given mass
region when $G_1$ and $G_2$ are properly chosen. Finally, the
strength $G_Q$ for quadrupole pairing was simply assumed to be
proportional to $G_M$, with a proportionality constant fixed in a
nucleus, choosing from the range 0.14 -- 0.18.

\subsection{The $^{178}$Hf example}

The nucleus $^{178}$Hf has become a discussion focus because of the
possibility to trigger the 2.45MeV, 31-year $16^+$-isomer decay. The
triggering could be made by applying external electromagnetic
radiation which, if successful, will lead to the controlled release
of nuclear energy \cite{Walker2}. Information on the detailed
structure as well as the transition of this and the surrounding
states thus becomes a crucial issue.  In the PSM calculation for
$^{178}$Hf \cite{Hf178}, the model basis was built with the
deformation parameters $\varepsilon_2=0.251$ and
$\varepsilon_4=0.056$.  Fig. 1 shows the calculated energy levels in
$^{178}$Hf, which are compared with the known data \cite{Mu97}.
Satisfactory agreement is achieved for most of the states, except
that for the bandhead of the first $8^-$ band and the $14^-$ band,
the theoretical values are too low.

It was found that the obtained states are generally $K$-mixed. If
the mixing is not strong, one may still talk about the dominant
structure of each band by studying the wavefunctions.  We found that
the $6^+$ band has mainly a 2-qp structure $\{ \nu [512]{5/2}^-
\oplus \nu [514]{7/2}^- \},$ the $16^+$ band has a 4-qp structure
$\{ \nu [514]{7/2}^- \oplus \nu [624]{9/2}^+ \oplus \pi [404]{7/2}^+
\oplus \pi [514]{9/2}^- \},$ the first (lower) $8^-$ band has a 2-qp
structure $\{ \nu [514]{7/2}^- \oplus \nu [624]{9/2}^+ \},$ the
second (higher) $8^-$ band has a 2-qp structure $\{ \pi [404]{7/2}^+
\oplus \pi [514]{9/2}^- \},$ and the $14^-$ band has a 4-qp
structure $\{ \nu [512]{5/2}^- \oplus \nu [514]{7/2}^- \oplus \pi
[404]{7/2}^+ \oplus \pi [514]{9/2}^- \}.$   These states, together
with many other states (not shown in Fig. 1) obtained from a single
diagonalization, form a complete spectrum including the high-$K$
isomeric states.

As far as energy levels are concerned, the PSM can give a reasonable
description simultaneously for multiple bands. The next question is
how electromagnetic transitions are described. The electromagnetic
transition between any two of these states can be directly
calculated \cite{PSM3} by using the wavefunctions. This is a crucial
test for the model wavefunctions.

\begin{figure}
\includegraphics[height=.25\textheight]{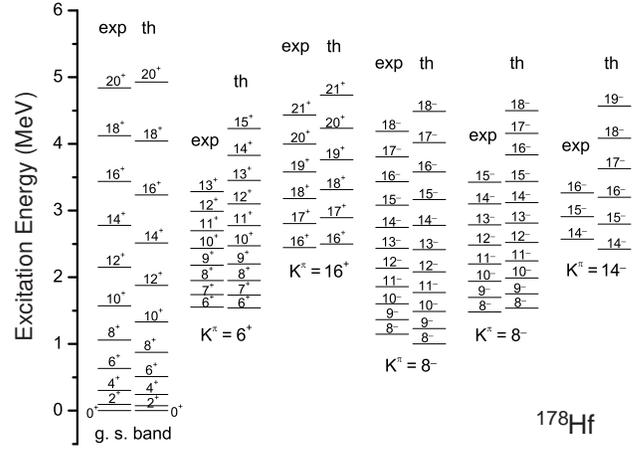}
\caption{Comparison of the PSM calculation with data for the
rotational bands in $^{178}$Hf .  This figure is adopted from Ref.
\cite{Hf178}.}
\end{figure}

\subsection{The $N=104$ isotones}

There have been detailed experimental studies on the $6^+$ isomer in
some $N=104$ isotones \cite{Yb174,Hf176,W178}. These data show that
along the $N=104$ isotones, lifetime of the $6^+$ isomer can vary
very much, differing by several orders of magnitude. Understanding
the underlying physics is a challenging problem: what is the
microscopic mechanism for such a drastic change in the neighboring
isotones?

\begin{table}
\caption{Comparison of calculated $^{174}$Yb ground band with data.
$E(I)$ are in keV and $B(E2, I\rightarrow I-2)$ in W.u..}
\label{tab:1}
\begin{tabular}{c|c|c|c|c}
\hline\noalign{\smallskip}
Spin $I$ & $E(I)$, Exp & $E(I)$, PSM & $B(E2)$, Exp & $B(E2)$, PSM \\
\noalign{\smallskip}\hline\noalign{\smallskip}
2 & 76.5  & 71.1  & 201(7)  & 195.18  \\
4 & 253.1 & 236.7 & 280(9)  & 279.01  \\
6 & 526.0 & 496.1 & 370(50) & 307.59  \\
8 & 889.9 & 848.0 & 388(21) & 322.31  \\
10& 1336  & 1290.3 & 325(22) & 331.28 \\
12& 1861  & 1820  & 369(23) & 337.13  \\
14& 2457  & 2433  & 320     & 340.91  \\
\noalign{\smallskip}\hline
\end{tabular}
\end{table}

\begin{table}
\caption{Comparison of calculated $^{174}$Yb $6^+$ isomer band with
data. $E(I)$ are in keV and $B(E2, I\rightarrow I-2)$ in W.u..}
\label{tab:2}
\begin{tabular}{c|c|c|c}
\hline\noalign{\smallskip}
Spin $I$ & $E(I)$, Exp & $E(I)$, PSM & $B(E2)$, PSM \\
\noalign{\smallskip}\hline\noalign{\smallskip}
6 & 1518.0 & 1503 &  \\
7 & 1671.1 & 1683 &  \\
8 & 1844.7 & 1886 & 36.76 \\
9 & 2038.3 & 2117 & 78.18 \\
10& 2251.5 & 2372 & 115.81 \\
11& 2483.7 & 2652 & 147.96 \\
12& 2734.4 & 2956 & 174.91 \\
13& 3003.1 & 3283 & 197.41 \\
\noalign{\smallskip}\hline
\end{tabular}
\end{table}

\begin{table}
\caption{Comparison of calculated inter-band transition of
$^{174}$Yb $6^+$ isomer to ground band. $B(E2)$ is in $e^2 fm^4$}
\label{tab:3}
\begin{tabular}{c|c|c}
\hline\noalign{\smallskip}
Transition & $B(E2)$, Exp & $B(E2)$, PSM \\
\noalign{\smallskip}\hline\noalign{\smallskip}
$6_{i}\rightarrow 4_{g}$ & $4.3(8)\times 10^{-9}$ & $8.49\times 10^{-8}$ \\
\noalign{\smallskip}\hline
\end{tabular}
\end{table}

PSM calculations are performed for $^{174}$Yb. The deformed basis is
constructed with deformation parameters $\varepsilon_2=0.275$ and
$\varepsilon_4=0.042$. In Tables I, II, and III, three groups of
results are listed, for the $K=0$ ground band (Table I) and $K=6$
isomer band (Table II) with in-band transitions, and inter-band
transitions (Table III) between the ground band and the $K=6$ isomer
band. These results suggest that while the energy levels for both
ground and isomer bands are reproduced, the $E(2)$ transition
probabilities are also correctly obtained. In particular, the
calculation yields a reasonable value of the very small inter-band
$B(E2)$ as what was observed in $^{174}$Yb \cite{Yb174} (see Table
III). Note that without mixing configurations in the wavefunction, a
direct transition from the $6^+$ isomer to ground band would be
forbidden. The obtained amount of inter-band $B(E2)$, though small,
is the consequence of $K$-mixing contained in the PSM.

On the other hand, in its isotones $^{176}$Hf and $^{178}$W, much
enhanced $B(E2)$ from the $6^+$ isomer to ground band has been
obtained experimentally. The values are $1.8\times 10^{-5}$
e$^2$fm$^4$ for $^{176}$Hf and $2.6\times 10^{-2}$ e$^2$fm$^4$ for
$^{178}$W. If a PSM calculation is performed for these two isotones,
one gets similar small numbers for the inter-band $B(E2)$ as in
$^{174}$Yb, which disagree with data. We have to conclude that
although the current PSM has $K$-mixing mechanism in the model,
which effectively introduces $\gamma$, the mixing within the
truncated space is apparently too weak.

\begin{figure}
\includegraphics[height=.23\textheight]{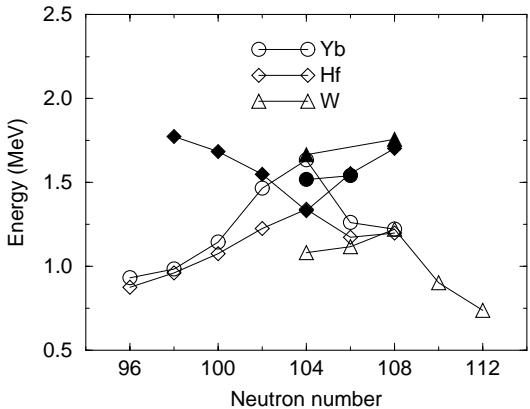}
\caption{Experimental data for excitation energy of $I^\pi=6^+$
isomer (filled symbols) and $2^+$ $\gamma$ vibrational state (open
symbols).}
\end{figure}

In Fig. 2, we plot experimental excitation energies of the $6^+$
isomer states together with $2^+$ state of $\gamma$ vibration for
Yb, Hf, and W isotopes. There seems to be an correlation between the
two plotted quantities. The correlation is such that to compare with
the $2^+$ $\gamma$ states, energy of the $6^+$ isomers shows an
opposite variation trend with neutron number. At $N=104$, nuclei
have the highest excitation of $\gamma$ vibrational states while
they show a minimum in $6^+$ isomer energy. In Fig. 2, $^{174}$Yb
appears to be the only nucleus in the collection that has the $6^+$
isomer lower than the $\gamma$ states. Therefore, below the $6^+$
isomer in $^{174}$Yb, there are no $\gamma$ states carrying finite
$K$ to be mixed in the wavefunction. This may have naively explained
why the $^{174}$Yb isomer decay is so exceptionally hindered.

In Ref. \cite{Shimizu}, a $\gamma$-tunneling model was introduced by
Narimatsu, Shimizu, and Shizuma to describe the enhanced $B(E2)$
values. In their model, the $\gamma$ degree of freedom is taken into
account, which breaks the axial symmetry explicitly. The spontaneous
symmetry breaking helps in realizing larger electromagnetic
transitions which would otherwise be impossible due to the selection
rule. In this way, the authors in \cite{Shimizu} were able to
describe the observed large inter-band $B(E2)$ in $^{176}$Hf and
$^{178}$W rather successfully. However, their model could not give
the above-discussed small $6^+_i\rightarrow 4^+_g$ inter-band
$B(E2)$ in $^{174}$Yb.

Both methods, the configuration mixing implemented by the PSM and
the $\gamma$-tunneling by Narimatsu {\it et al.}, introduce a
mechanism to break the axial symmetry; however the degree of
symmetry breaking is different. If the physical process is a
perturbation in the $K$ space, then it is better described by the
PSM based on the axially symmetric mean field. If it is not, axial
symmetry in the mean field must be broken as in the
$\gamma$-tunneling model. The two models may be viewed as two
different simplifications of the complicated many-body problem; each
emphasizes one aspect. It is desired that one can have one unified
microscopic description for all cases.

To efficiently introduce $\gamma$ degree of freedom within the PSM,
one can break the axial symmetry of the single-particle basis and
carry out three-dimensional angular momentum projection. The shell
model diagonalization is then performed in the projected
multi-quasiparticle configurations based on $\gamma$ deformed basis.
One example is the description of $\gamma$ vibrational states within
the PSM. It was shown \cite{Sun00} that by using projected
triaxially-deformed basis, it is possible to describe the ground
band and $\gamma$ band simultaneously. Thus, an extended PSM that
introduces triaxiality in the model would be useful for cases with
large $K$-violation. Such an extension has recently be developed for
odd-odd nuclei \cite{Gao06a} and even-even nuclei \cite{Javid08},
and will be applied to the isomer study.

\section{Shape isomer and configuration mixing}

Coexistence of two or more well-developed shapes at comparable
excitation energies is a well-known phenomenon. The expected nuclear
shapes include, among others, prolate and oblate deformations. In
even-even nuclei, an excited 0$^+$ state may decay to the ground
0$^+$ state via an electric monopole (E0) transition. For lower
excitation energies, the E0 transition is usually very slow, and
thus the excited 0$^+$ state becomes a shape isomer.

\subsection{Shape isomer in $^{68}$Se and $^{72}$Kr and the impact
on isotopic abundance in X-ray bursts}

\begin{figure}
\includegraphics*[angle=0,width=20pc]{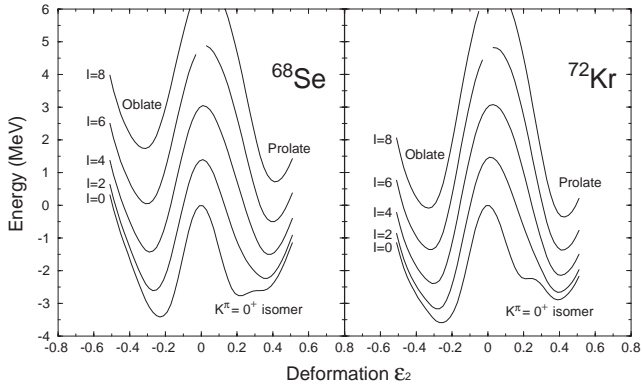}
\caption{Energy surfaces for various spin states in $^{68}$Se and
$^{72}$Kr as functions of deformation variable $\varepsilon_2$. This
figure is adopted from Ref. \cite{Wiescher05}.}
\end{figure}

Fig. 3 shows calculated projected energies as a function of
deformation variable $\varepsilon_2$ for different spin states in
the $N=Z$ nuclei $^{68}$Se and $^{72}$Kr. The configuration space
and the interaction strengths in the Hamiltonian are the same as
those employed in the previous calculations for the same mass region
\cite{Sun04}. It is found that in both nuclei, the ground state
takes an oblate shape with $\varepsilon_2\approx -0.25$. As spin
increases, the oblate minimum moves gradually to
$\varepsilon_2\approx -0.3$. Another local minimum with a prolate
shape ($\varepsilon_2\approx 0.4$) is found to be 1.1 MeV
($^{68}$Se) and 0.7 MeV ($^{72}$Kr) high in excitation. Bouchez {\it
et al.} \cite{Kr72} observed the 671 keV shape-isomer in $^{72}$Kr
with half-life $\tau = 38\pm 3$ ns. The one in $^{68}$Se is the
prediction, awaiting experimental confirmation. Isomers in these
nuclei have also been predicted by Kaneko, Hasegawa, and Mizusaki
\cite{Kaneko04}.

The existence of low energy 0$^+$ shape isomer along the $N=Z$
nuclei has opened new possibilities for the rp-process
\cite{rpReport} reaction path occurring in X-ray burst. Since the
ground states of $^{73}$Rb and $^{69}$Br are bound with respect to
these isomers, proton capture on these isomers may lead to
additional strong feeding of the $^{73}$Rb($p,\gamma$)$^{74}$Sr and
$^{69}$Br($p,\gamma$)$^{70}$Kr reactions. However, the lifetime of
the isomeric states must be sufficiently long to allow proton
capture to take place. No information is available about the
lifetime of the $^{68}$Se isomer while the 55 ns lifetime of the
isomer in $^{72}$Kr is reported \cite{Kr72}. Based on Hauser
Feshbach estimates \cite{rpReport} the lifetime against proton
capture is in the range of $\approx$100 ns to 10 $\mu$s depending on
the density in the environment. Considering the uncertainties in the
present estimates a fair fraction may be leaking out of the
$^{68}$Se, $^{72}$Kr equilibrium abundances towards higher masses.

\subsection{Configuration mixing with different shapes}

To calculate isomer lifetime, decay probability is needed. This
involves transitions from the shape isomer to the ground state,
which belong to different shapes or deformation minima. If the
energy barrier between the minima is not very high, configuration
mixing of the two shapes must be taken into account. In the
following, we outline a scheme to consider such a mixing. The
discussion is general; the shapes can be any kinds of two deformed
ones in a nucleus. For example, one of them can be a
prolately-deformed and the other an oblately-deformed shape, or one
of them can be a normally-deformed and the other a superdeformed
shape. Generalizing the method further, it can describe those
transitional nuclei where energy surfaces are typically flat.

The heart of the present consideration is the evaluation of
overlapping matrix element in the angular-momentum-projected bases.
Let us start with the PSM wave function in Eq. (\ref{wavef})
\begin{equation}
|\psi^{I,\sigma}_M\rangle = \sum_{\kappa} f^{I,\sigma}_{\kappa} \hat
P^{\,I}_{MK_\kappa}|\phi_\kappa\rangle . \nonumber
\end{equation}
For an overlapping matrix element, states in the left and right hand
side must correspond to different deformed shapes. Therefore, two
different sets of quasiparticle generated at different deformations
are generally involved. Let us denote $\left|\phi_\kappa\right>$
explicitly as $\left|\phi_\kappa(a)\right>$ and
$\left|\phi_\kappa(b)\right>$, for which we define two sets of
quasiparticle operators $\{a^\dagger\}$ and $\{b^\dagger\}$
associated with the quasiparticle vacua $\left|a\right>$ and
$\left|b\right>$, respectively.

For simplicity, we assume axial symmetry. The general
three-dimensional angular momentum projection is reduced to a
problem of one-dimensional projection, with the projector having the
following form
\begin{equation}
\hat P^I_{MK} = \left(I+{1\over 2}\right) \int^\pi_0 d\beta ~~{\rm
sin} \beta ~~d^I_{MK}(\beta) ~~\hat R_y(\beta) \label{proj}
\end{equation}
with
\begin{equation}
\hat R_y(\beta) = e^{-i\beta \hat J_y}.
\end{equation}
In Eq. (\ref{proj}), $d^I_{MK}(\beta)$ is the small-$d$ function and
$\beta$ is one of the Euler angels. The evaluation of the
overlapping matrix element is eventually reduced to the problem
\begin{equation}
\left<\Phi_{\kappa'}(b)\right| \hat O \hat R_y(\beta)
\left|\Phi_{\kappa}(a)\right>, \label{RMT}
\end{equation}
which is the problem of calculating the $\hat O$ operator sandwiched
by a multi-qp state $\left| \Phi_{\kappa'}(b)\right>$ and a {\it
rotated} multi-qp state $\hat R_y(\beta) \left|
\Phi_{\kappa}(a)\right>$, with $a$ and $b$ characterizing different
qp sets. In Eq. (\ref{RMT}), $\hat O$ stands for $\hat H$ or 1.

To calculate $\left<\Phi_{\kappa'}(b)\right| \hat O \hat R_y(\beta)
\left|\Phi_{\kappa}(a)\right>$, one must compute the following types
of contractions for the Fermion operators
\begin{eqnarray}
A_{ij} &=& \left<b\right| [\beta] a^\dagger_ia^\dagger_j
 \left|a\right> = [V(\beta)U^{-1}(\beta)]_{ij},
\nonumber\\
B_{ij} &=& \left<b\right|b_ib_j [\beta] \left|a\right> =
 [U^{-1}(\beta)V(\beta)]_{ij},
\label{ABCmatrix}\\
C_{ij} &=& \left<b\right|b_i [\beta] a^\dagger_j \left|a\right> =
[U^{-1}(\beta)]_{ij} , \nonumber
\end{eqnarray}
where we have defined
\begin{equation}
[\beta]={{\hat R_y(\beta)}\over {\left<b\right| \hat R_y(\beta)
\left|a\right>}}, \nonumber
\end{equation}
and
\begin{equation}
\left<b\right|\hat R_y(\beta)\left|a\right> = [{\rm det}~
U(\beta)]^{1/2}. \label{DET}
\end{equation}
Eqs. (\ref{ABCmatrix}) and (\ref{DET}) are written in a compact form
of $N\times N$ matrix, with $N$ being the number of total single
particles. The general principle of finding $U(\beta)$ and
$V(\beta)$ is given by the Thouless theorem \cite{Thouless60}, and a
well worked-out scheme can be found in the work of Tanabe {\it et
al.} \cite{Tanabe99} (see also Ref. \cite{Gao06b}).

To write the matrices $U(\beta)$ and $V(\beta)$ explicitly, we
consider the fact that $\{a_i, a^\dagger_i\}$ and $\{b_i,
b^\dagger_i\}$ can both be expressed by the spherical representation
$\{c_i,c^\dagger_i\}$ through the HFB transformation
\begin{equation}
\begin{array}{rcl}
\left[ \begin{array}{c} c \\ c^\dagger
\end{array} \right]
&=& \left( \begin{array}{cc} U_a & V_a \\ V_a & U_a
\end{array} \right)
\left[ \begin{array}{c} a \\ a^\dagger
\end{array} \right]
\vspace{8pt}\\
\left[ \begin{array}{c} c \\ c^\dagger
\end{array} \right]
&=& \left( \begin{array}{cc} U_b & V_b \\ V_b & U_b
\end{array} \right)
\left[ \begin{array}{c} b \\ b^\dagger
\end{array} \right].
\label{abcd}
\end{array}
\end{equation}
$U_a, V_a, U_b$ and $V_b$ in above equations, which define the HFB
transformation, are obtained from the Nilsson-BCS calculation. A
rotation of the spherical basis can be written in a matrix form as
\begin{eqnarray}
\hat R_y(\beta) \left[ \begin{array}{c} c \\ c^\dagger
\end{array} \right]
\hat R_y^\dagger (\beta) &=& \left( \begin{array}{cc} d(\beta) & 0
\\ 0 & d(\beta)
\end{array} \right)
\left[ \begin{array}{c} c \\ c^\dagger
\end{array} \right].
\label{transf}
\end{eqnarray}
Combining Eqs. (\ref{abcd}) and (\ref{transf}) and noting the
unitarity of the HFB transformation, one obtains
\begin{eqnarray}
\hat R_y(\beta) \left[ \begin{array}{c} b \\ b^\dagger
\end{array} \right]
\hat R_y^\dagger (\beta) &=& \left( \begin{array}{cc} U_b & V_b \\
V_b & U_b
\end{array} \right)^T
\left( \begin{array}{cc} d(\beta) & 0 \\ 0 & d(\beta)
\end{array} \right) \nonumber\\ &\times&
\left( \begin{array}{cc} U_a & V_a \\ V_a & U_a
\end{array} \right)
\left[ \begin{array}{c} a \\ a^\dagger
\end{array} \right].
\end{eqnarray}
$U(\beta)$ and $V(\beta)$ can finally be obtained from the following
equation
\begin{eqnarray}
\begin{array}{rcl}
&\left( \begin{array}{cc} U(\beta)&V(\beta)\\V(\beta)&U(\beta)
\end{array} \right)
= \left( \begin{array}{cc} U^T_b & V^T_b \\ V^T_b & U^T_b
\end{array} \right)
\left( \begin{array}{cc} d(\beta) & 0 \\ 0 & d(\beta)
\end{array} \right)
\left( \begin{array}{cc} U_a & V_a \\ V_a & U_a
\end{array} \right)
\vspace{10pt}\nonumber\\
&= \left( \begin{array}{cc} U^T_bd(\beta)U_a+V^T_bd(\beta)V_a &
U^T_bd(\beta)V_a+V^T_bd(\beta)U_a
\vspace{6pt}\\
V^T_bd(\beta)U_a+U^T_bd(\beta)V_a &
U^T_bd(\beta)U_a+V^T_bd(\beta)V_a
\end{array} \right).
\end{array}
\end{eqnarray}

With the overlapping matrix elements that connect configurations
belonging to different shapes, one obtains wavefunctions containing
configuration mixing. Using these wavefunctions, one can further
calculate inter-transition probabilities from a shape isomer to the
ground state.

\section{Summary}

We have introduced the Projected Shell Model description for two
kinds of isomers, $K$-isomers and shape isomers. We have shown that
the physics of $K$-mixing in multi-qp states is well incorporated in
the model with the basis states having axial symmetry.
Diagonalization mixes configurations of different $K$, which
effectively introduces triaxiality. For $K$-isomers with much
enhanced decay probability to the ground state, a triaxial PSM is
needed, which employs $\gamma$ deformed basis states. On the other
hand, projected energy surface calculations have led to a picture of
shape coexistence. A scheme has been developed which allows
calculations for transition between a shape isomer and the ground
state.

The author is grateful to J. Hirsch and V. Velazquez for warm
hospitality during the Cocoyoc 2008 meeting. He acknowledges helpful
discussions with A. Aprahamian, Y. R. Shimizu, P. M. Walker, and M.
Wiescher. This work is supported by the Chinese Major State Basic
Research Development Program through grant 2007CB815005 and by the
U. S. National Science Foundation through grant PHY-0216783.

\end{document}